\documentclass{iau} 
\usepackage{mwe}    % loads »blindtext« and »graphicx«
\usepackage{subfig}
\usepackage{hyperref}

\title[Bridging Galaxy Dynamics and Baryon Efficiency] %% give here short title %%
{Bridging Galaxy Dynamics and Baryon Efficiency of 40 EDGE-CALIFA galaxies}

\author[Kalinova et al.]   %% give here short author list %%
{Veselina Kalinova$^1$, Dario Colombo$^1$, Erik Rosolowsky$^1$, and the EDGE-CALIFA collaboration}

\affiliation{$^1$Department of Physics 4-181 CCIS, University of Alberta, 
Edmonton AB T6G2E1, Canada \\ email: {\tt veselina@ualberta.ca} \\%[\affilskip]
}

\pubyear{2015}
\volume{315}  %% insert here IAU Symposium No.
\jname{From interstellar clouds to star-forming galaxies: \\ universal processes?}
\editors{A.C. Editor, B.D. Editor \& C.E. Editor, eds.}
\begin{document}

\maketitle

\begin{abstract}
We apply the Jeans Axisymmetric Multi-Gaussian Expansion method to the stellar kinematic maps of 40 Sa--Sd EDGE-CALIFA galaxies and derive their circular velocity curves (CVCs). The CVCs are classified using the Dynamical Classification method developed in \cite[Kalinova et al. (2015)]{kalinova15}. We also calculate the observational baryon efficiency, OBE, where $M_*/M_b=M_*/(M_*+M_{HI}+M_{H_2})$ of the galaxies using their stellar mass, total neutral hydrogen mass and total molecular gas from CO luminosities. Slow-rising, Flat and Round-peaked CVC types correspond to specific OBEs, stellar and dark matter (DM) halo mass values, while the Sharp-peaked CVCs span in the whole DM halo mass range of $10^{11}-10^{14}M_{\odot}$.

%% add here a maximum of 10 keywords, to be taken form the file <Keywords.txt>
\end{abstract}

%\firstsection % if your document starts with a section,
              % remove some space above using this command.

%%%%% Introduction

{\bf Introduction.} Galaxy formation and evolution are regulated by the interaction between three fundamental components: stars, gas, and dark matter. To understand the link between these components, modern simulations have used the ``abundance matching'' where empirical relationships between stellar mass, star formation (SF) efficiencies, and the DM halo mass ($M_h$) are calibrated using populations of galaxies at different redshifts (\cite[Moster et al. 2013]{moster13}). This model proposes two domains based on SF feedback (low-mass) vs. Active Galactic Nuclei feedback (high-mass) tied to specific values of $M_h$ and baryon efficiency. The parameters in the relationship imply globally important physics as fundamental as the Kennicutt-Schmidt law (\cite[Kennicutt et al. 1998]{kennicutt98}), and the origin of the cold gas phase with respect to the stellar disc (\cite[Davis et al. 2011]{davis11}).

%%%%%% Data and Method 

{\bf Data and Method.} To observationally test the theoretical prediction of \cite[Moster et al. (2013)]{moster13} we focused on 40 Sa--Sd galaxies from the optical integral-field spectroscopic survey CALIFA\footnote{\url{http://califa.caha.es}} (\cite[S{\'a}nchez et 
al. 2012]{sanchez12}) and the CARMA-CO survey EDGE\footnote{\url{http://www.astro.umd.edu/~bolatto/EDGE/}} (Bolatto et al., in prep). 
Dynamical models and CVCs of the galaxies are constructed using the Multi-Gaussian Expansion method (MGE; \cite[Emsellem et al. 1994]{emsellem94}) on SDSS images and Jeans Anisotropic MGE model (JAM; \cite[Cappellari 2008]{cappellari08}) on CALIFA stellar kinematics (Falc{\'o}n-Barroso et al., in prep). Further, we classify the CVCs of the 40 galaxies by shape and amplitude using the Dynamical classification method, described in \cite[Kalinova et al. 2015]{kalinova15}. The principal components of our sample are obtained through the eigenvectors of the prototype curves in Figure 7 of their paper, i.e. a given curve is associated to a specific dynamical class (Slow-rising, Flat, Sharp-peaked, Round-peaked) based on the minimum Euclidean distance to the k-means centroids that define the prototype curves in the five-dimensional principal components clustering space. The final dynamical classification of our 40 galaxies is shown in Fig.1a. To understand how these four dynamical classes are related to the OBE and $M_h$ of the galaxies we use also measurements of the total stellar mass ($M_*$) adopted from \cite[Gonz{\'a}lez-Delgado et al. (2015)]{gonzalez15}, while the $M_{HI}$ are taken from \cite[Springob et al. (2005)]{springob05}. DM halo mass estimations are extrapolated from the $M_*-M_h$ relation of \cite[Moster et al. (2013)]{moster13} using our adopted stellar masses (see Fig.1b, top).  The CO luminosity of the molecular gas is derived from the CARMA CO(1-0) EDGE integrated intensity maps (Bolatto et al., in prep), and then converted in total $M_{H_2}$ using the factor $\alpha_{CO}=$4.4 $M_{\odot}$ $\mathrm{(pc^{2}}$ K km $s^{-1})^{-1}$.

%Figures

\begin{figure}[!ht]
    \subfloat[\label{subfig-1:dummy}]{%
      \includegraphics[width=0.65\textwidth]{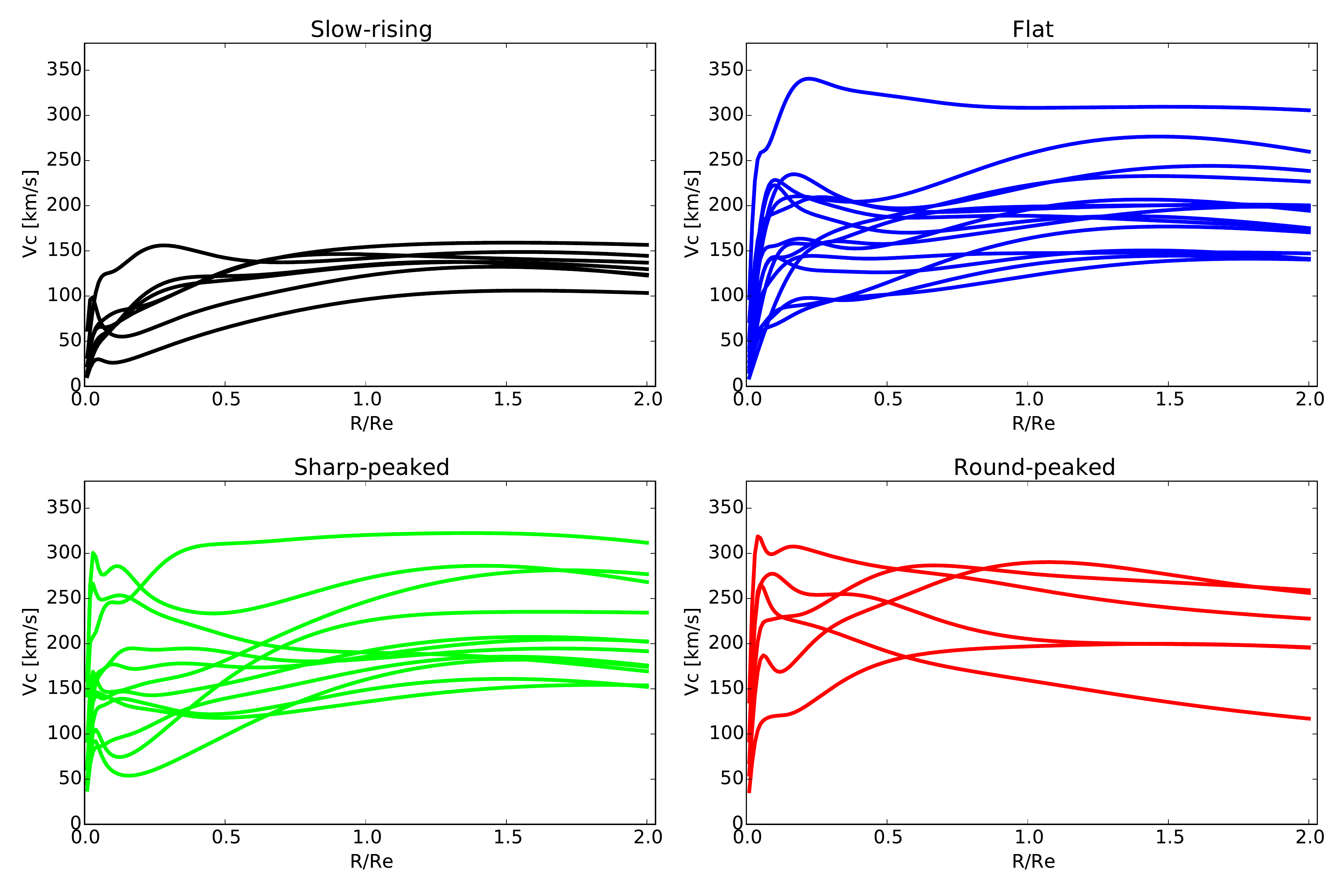}
    }
    \hfill
    \subfloat[\label{subfig-2:dummy}]{%
      \includegraphics[width=0.33\textwidth]{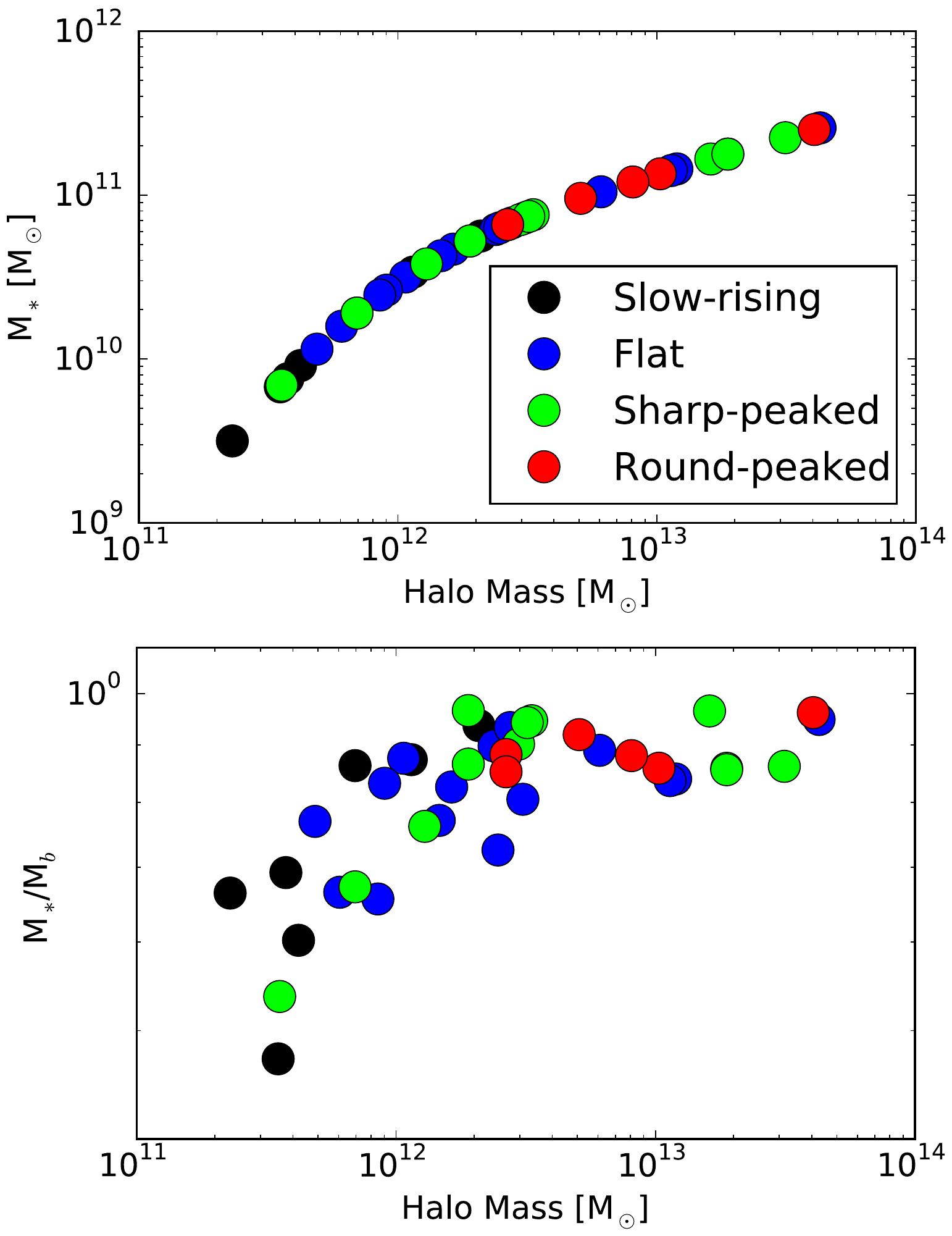}
    }
    \caption{ {\footnotesize{\bf (a)} Dynamical Classification of 40 (Sa--Sd) EDGE-CALIFA galaxies.
   {\bf(b)}{\it Top:} Extrapolated DM Halo masses from the $M_*-M_h$ relation of \cite[Moster et al. (2013)]{moster13} using the stellar mass of the galaxies from \cite[Gonz{\'a}lez-Delgado et al. (2015)]{gonzalez15}. {\it Bottom:} Observational Baryon Effeciencies ($M_*/M_b$) of the galaxies throughout their dynamical classes, shown in different color.
   }}
    \label{fig:dummy}
  \end{figure}
%
%%%% Results
{\bf Results.} Our analysis shows that Slow-rising CVCs are manly presented by low $M_h$ ($10^{11}-10^{12} M_{\odot}$) and low observational baryon efficiencies, while Round-peaked CVCs correspond to galaxies with high OBE and high implied $M_h$ ($10^{12}-10^{14} M_{\odot}$). Flat CVCs cluster around intermediate DM halos, i.e., $M_h$ ($10^{12}-10^{13} M_{\odot}$) with approximately constant and high OBEs. Sharp-peaked CVCs, instead, span the entire OBE range and DM halo mass range ($10^{11}-10^{14}M_{\odot}$, see Fig 1b). The dynamics of the 40 galaxies, and in particular, the shape and the amplitude of their CVCs might be linked to the cooling of the hot gas and formation of the stellar disc in the different DM halos mass ranges. This represents the key to unveil the connection between internal and external processes in the galaxies. Better conclusions will be drawn via DM halo masses measured by high resolution HI rotation curves and modelling of the hot gas content.

%=====================================================

\end{document}